# Implementation of a 300 kA Pulsed Power Supply

C.C. Jensen, N. Gurley, H. Pfeffer
Fermi National Accelerator Lab
Batavia IL USA
ccjensen@fnal.gov, ngurley@fnal.gov, pfeffer@fnal.gov

***Abstract*— The Long Baseline Neutrino Facility (LBNF) will produce the world's most intense neutrino beam. While the magnetic horns are expected to be replaced, the LBNF horn power supply is expected to last the lifetime of the project, 30 years. Three series connected magnetic horns will require a 300kA, 800µs pulses at a rate of 1.4 pps to focus the beam. Prototyping proved the component choices and designs. Test results for a single cell and the full 32 cell system will be presented.**

## I. Introduction

The design and design tradeoffs for making this pulsed supply were previously discussed [1]. The supply consists of 32 pulsed power cells; identical capacitor banks, thyristor switches, diodes and several inductors, Fig 1. All the cells discharge in parallel through a 9 layer stripline [2] with the outside two layers conducting half as much current as the other layers. All the stripline layers connect at the horn making a half sine wave current pulse defined by the pulser total capacitance and the inductance of the horn and stripline. The recovery choke then rings the pulse capacitor back to the original polarity. This paper focus on prototype testing and issues with construction.

Realization of the supply started with prototype procurement and testing of a single cell. Two very similar pulsed supplies are needed, one for testing individual horns and the operational one for running the three horns in series. The only difference between the supplies is the operating voltage, energy storage capacitance (keeping the pulse width the same) and thyristor snubber network: 2.5 kV, 32 cells x 1.9 mF and a simple R-C snubber for the horn test supply; 5 kV, 32 cells x 0.6 mF and an R-C-D snubber for the operational supply. There are 16 enclosures each containing two independent cells. Each enclosure is about 1.1 m x 1.8 m x 2.8 m tall and weighs 2300 kg fully assembled.

## II. Prototype Testing

First testing was voltage sharing of the four series thyristors, Fig 1 C-D. Turn on and turn off sharing were measured during normal operation with a load of 100 µH, equivalent to the 32 times the nominal load inductance. The saturating inductor limited initial turn on and turn off dI/dt to ~10 A/us then increasing to ~30 A/us after saturation. The cell internal inductance, based on fitting a model to measurements, is 10 µH including connections to 9 layer stripline. The saturating inductance is estimated to be 4 µH. The saturating inductor transitions at ~ 500 A going into saturation and ~ 400 A coming out of saturation We also measured turn off voltage sharing with a short at the stripline connection, M to F in Fig 1. This is the main reason for the saturating choke. In this case we measured ~360 A/us after the choke saturated and ~20 A/us with the choke unsaturated. The thyristors have this lower dI/dt for ~60 µs before current reversal. The turn off voltage overshoot is ~10 % above nominal in this fault condition with the R-C-D snubber. The reduced dI/dt from the saturating choke reduces the $Q_{rr}$ and $I_{rr}$ to practically the nominal operation levels. Diode voltage sharing during turn on and turn off was also verified to be acceptable.

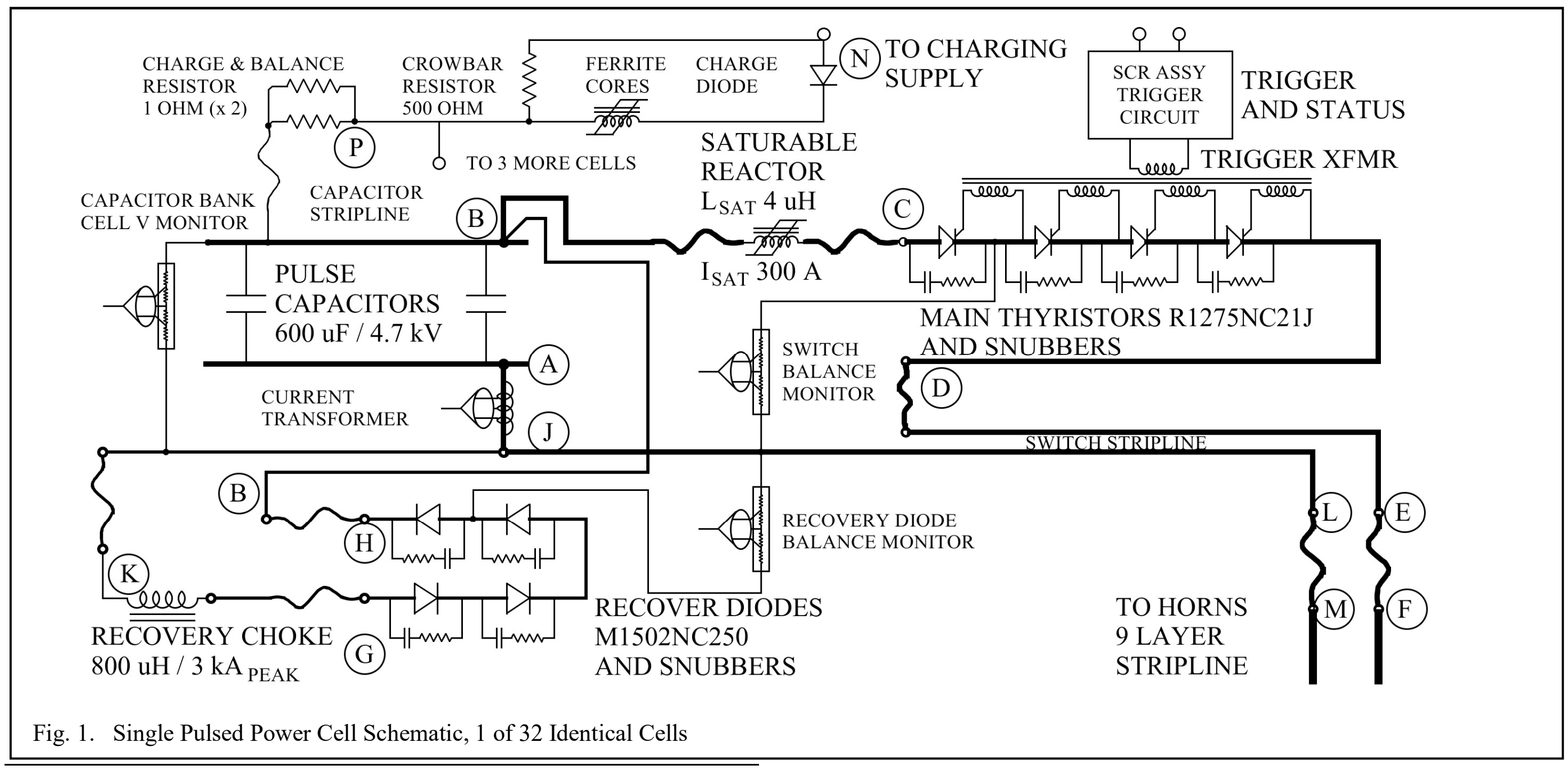

Fig. 1. Single Pulsed Power Cell Schematic, 1 of 32 Identical Cells

There were no major issues during prototype testing with the thyristor, diode, capacitor stripline, switch stripline or capacitor assemblies but there were some minor issues. The capacitor cases were not well grounded which caused some interesting but small ground currents. The capacitors are properly grounded in the series production. We did over torque a capacitor bushing which resulted in a capacitor internal short. The result of the breakdown uncovered a high current fault path through the voltage monitor which was corrected. The initial current transformer (CT) saturated during short circuit testing and did not recover. The requirements were stated correctly but missed during the manufacturer's review. A physically larger device with the proper A-s rating was then received and successfully tested.

The original choice for thyristor is a R1275NC21 with $Q_{rr}$ matching from IXYS/Littelfuse. An alternative thyristor ATF63S21T with $Q_{rr}$ matching from Poseico was bought and tested. The turn on and turn off sharing were measured to be similar for the two devices. The temperature rise of the snubbers, diodes and thyristors were measured after pulsing at full repetition rate and it was confirmed that no forced air cooling was required. The DC voltage sharing resistors across the diodes and thyristors were the hottest. At the end of prototype testing we had about 30 million pulses on each element except the energy recovery choke and saturating choke magnetics. The main issues were with these pulsed magnetic elements and with the pulsed high current wire connections.

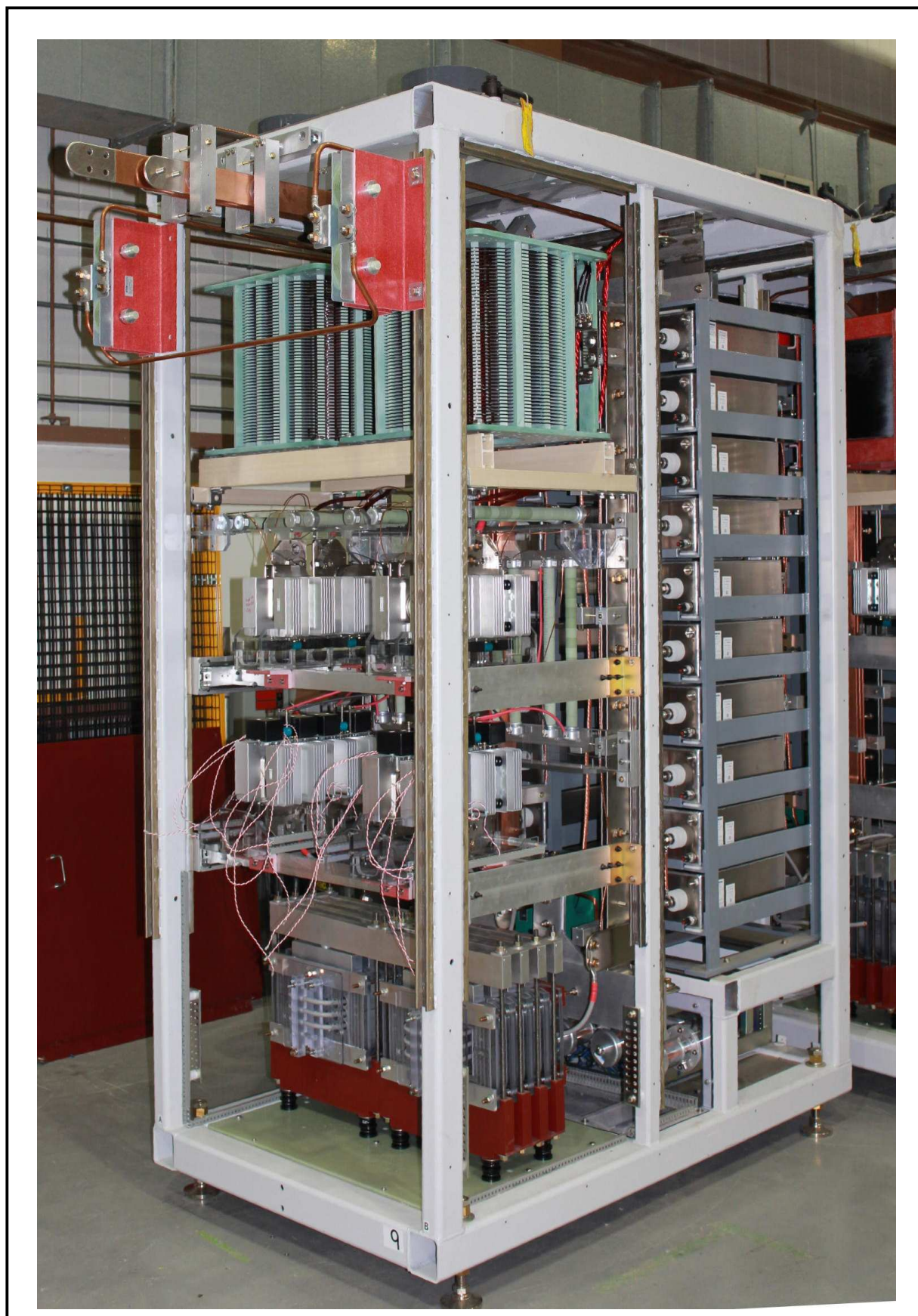

Fig. 2. Two Cell Enclosure, Test Supply Construction, 1 of 16 cabinets

The energy recovery choke is 800 uH with a peak current of 3 kA, a 2.3 ms half period, an rms current of 120 A and a peak voltage of 4 kV. Its purpose is to reverse the voltage polarity of the capacitor back to the initial state to reduce charging power. Total losses were specified to be <400 W at repetition rate 1.4 pps for a Q of approximately 13. A detailed specification was written and two vendors were chosen to make prototypes before the series production. Both constructions were single layer solenoid and filled the allowed space. One used Litz wire while the other used 3 parallel rectangular conductors. Both met the inductance, passed corona inception voltage test and met the loss requirement.

The Litz wire construction originally used a standard varnish to restrain the conductors. The conductors failed mechanically as the coil and conductors bounced around. A heavy varnish also failed. In addition the terminations of the wire failed mechanically but not thermally. The successful solution was to completely encapsulate the coil in epoxy and terminate the wire in a crimp lug. The parallel solid conductors failed mechanically as well. The initial bracing was made of GPO3 which broke and then the conductors cracked from repetitive stress. The bracing material was changed to G10, but the conductors still cracked. More substantial bracing for the coils was added finally. The wires cracked near the terminals on this choke as well until additional bracing was added. Success was defined as completion of 10M pulses at nominal current and repetition rate.

Both companies also made a 10 kA, 50 uH inductor, a load for testing individual cells. The Litz wire load mechanically failed quickly and was deemed unusable. The company that used rectangular conductors had bracing and wire fatigue failures. The changes in the design of the test load were the same as the recovery choke.

The saturating choke was designed in house since we had difficulty in getting vendors for the more standard energy recovery choke. The cores are 3% silicon steel, 0.004" lamination thickness tape wound, cut and lapped with extra epoxy reinforcement around the gap. The original design was for a 3 turn 8 core inductor with a .010" gap. Significant clamping force, 4000 lbs per core, was used to reduce motion and acoustic noise based on previous experience with saturating chokes. None of these efforts eliminated the noise or motion. The main acoustic noise came from vibration of the enclosure frame due to the impulse. Shock absorbing machine mounting feet were added to reduce coupling of energy into the frame and this helped some.

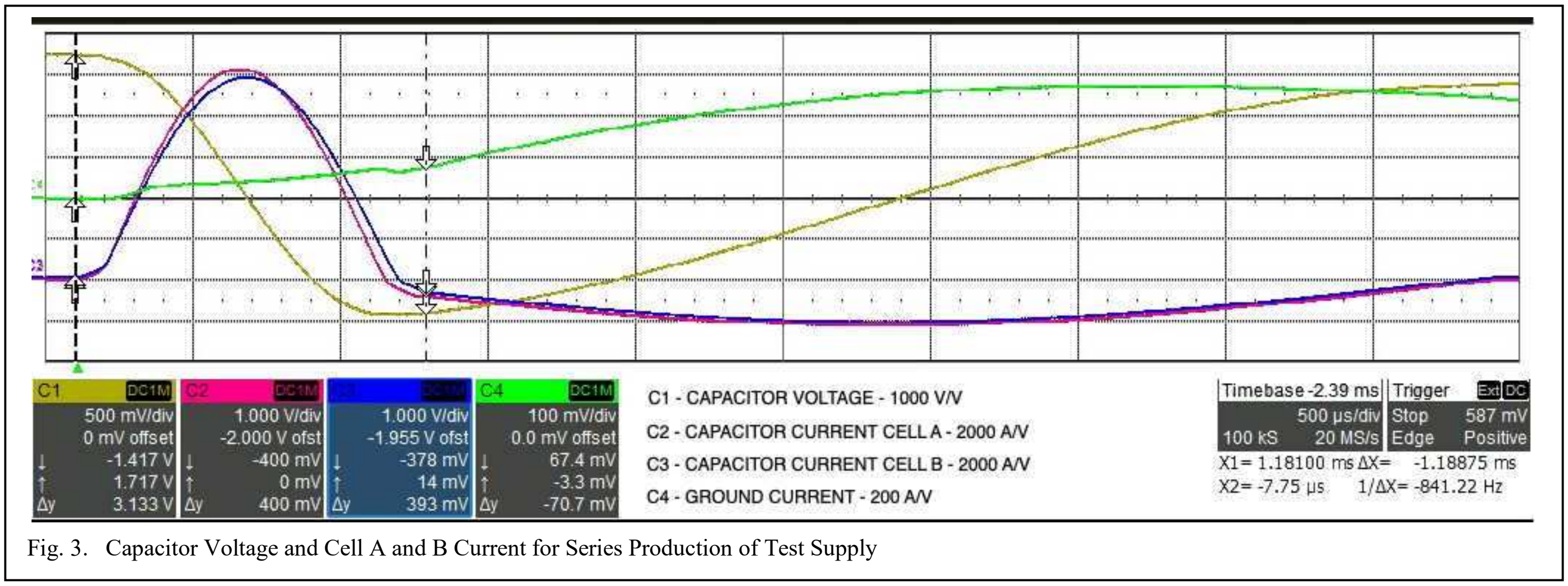

Fig. 3. Capacitor Voltage and Cell A and B Current for Series Production of Test Supply

The requirement was to saturate after 0.1 V-sec and have a saturated inductance of about 4 uH. This limits the short circuit fault current to 40 kA over the normal 10 kA. To reduce cost a 4 turn 6 core choke with a 0.015" gap was tested. The core window was based on clearances for 3 turns so adding the extra turn meant the winding geometry was more complicated. Litz wire, 80 strands of AWG 20, to reduce losses was used initially. Past experience is that smaller diameter wire will fail from cyclical stress due to metal fatigue. Several different methods of bracing and support of the windings were tried. The windings always failed mechanically, either at the crimp lug or in the winding. Any motion led to mechanical failure. We reverted back to 3 turn winding and used XHHW insulated AWG 1/0 wire with heavy duty heat shrink at the crimp lugs. Wire clamps on each end of the core were added to completely constrain the wire and a hard bus extension was added to reduce the needed length of the wire leads.

There is a requirement for reversing the polarity of the current in the load while also maintaining the lowest potential at the same location in the string of 3 horns. This means reversing the polarity of the charging supply and the diode and thyristor in each cell. The goal was to reverse all 32 cells in 8 hours. A flexible connection was chosen to avoid mechanically moving bus work or semiconductors as had been done in NuMI [3]. There are two sets of flexible leads needed to reverse polarity, one for the thyristor, Fig 1 C & D, and one for the diode, Fig 1 G & H. The third set connects the cell to the 9 layer stripline, Fig 1 L & E, but does not need to be moved during polarity changes. The number of switches is higher (32 switches of 10 kA each vs 12 switches of 17 kA each) while the peak current per switch is lower so a flexible wire seemed a feasible choice. The same Litz wire used in the saturating choke was chosen before the failures on it. There were multiple failures where the wires were crimped into a two hole lug after stripping off the varnish. The failures were due to repetitive motion which caused the small strands to break at the crimp. We also added support in various places to reduce wire motion during pulsing. We ultimately changed the termination method to a double hexagonal crimp and then a very stiff heat shrink over the barrel of the lug and the wire for additional support in addition to various wire supports along the length. However, this only passed the pulse test for the diode connections, which are 3 kA, and not the thyristor connections, which are 10 kA.

The 10 kA reversable connections were finally done with AWG 1/0 XHHW, 19 strands of 13 AWG. They are not very flexible, but the temperature rise is modest, 40 C. For the fixed 10 kA connections AWG 3/0 bare stranded wire was selected. The AWG 3/0 bare stranded wire mechanically failed until the same double hexagonal crimp and heavy duty heat shrink as used.

## III. Test Supply Series Construction

Once the test cell passed the 10 M pulse minimum life test at full current and repetition rate the final order for the recovery chokes and remaining saturating choke parts were placed. The decision was to keep both vendors for the recovery chokes. This keeps both vendors as an option for the operational supply and will allow further lifetime testing of the two constructions in the test supply.

The same number of series thyristors in a cell switch will be used for the test stand even though it is operating at half the voltage. A redesign would have meant the test system can not provide spares to the operational system. This also meant the thyristor snubber circuit could be modified to a simple RC networks for the test system since we have more voltage margin in the test stand. The snubber will be re-evaluated for the production system based on the test results from the prototype.

Both the thyristor and diode snubber component layout was simplified from point to point wire and bus to a PCB mounted directly to the high current bus work. This reduced assembly complexity. The test supply will have 31 of the 32 cells with the IXYS devices as well as one cell with the Poseico devices. While the 10 kA flexible lead issue could not be solved, the thyristor connections on this first supply do not reverse polarity. The test supply has one cell with a lead consisting of two Litz wire cables in parallel each with heavy strain relief to see how that will survive.

Each production unit was tested for at least 200,000 pulses at full peak current (10 kA peak) and repetition rate (1.43 pulses per second). The first few production units were tested for longer as assembly ramped up. These tests resulted in adding a few more wire clamps to reduce flexible lead motion further. Units were tested in one area and then moved to a different building where the test supply is assembled. This tested the ability to move the completed cabinets which is required for the final supply.

## IV. Stripline

This is a 9 conductor stripline, each conductor, 12" wide, 0.50" thick, 0.50" spacing between conductors. The conductors alternate supply and return current. Another requirement is to feed the horn in four places with currents that are equal within 1% to guarantee magnetic field uniformity in the horn. The stripline for the test system was assembled in 6 sections, the longest 4.4 m long. The test system has a total length of ~18 m while the operational system is ~72 m long. Most of the features that will exist in the operational system are included in the test system except an expansion joint and a three way tee. The individual sections were assembled on a bench and then hipotted successfully to 15 kV DC, three times the operational voltage.

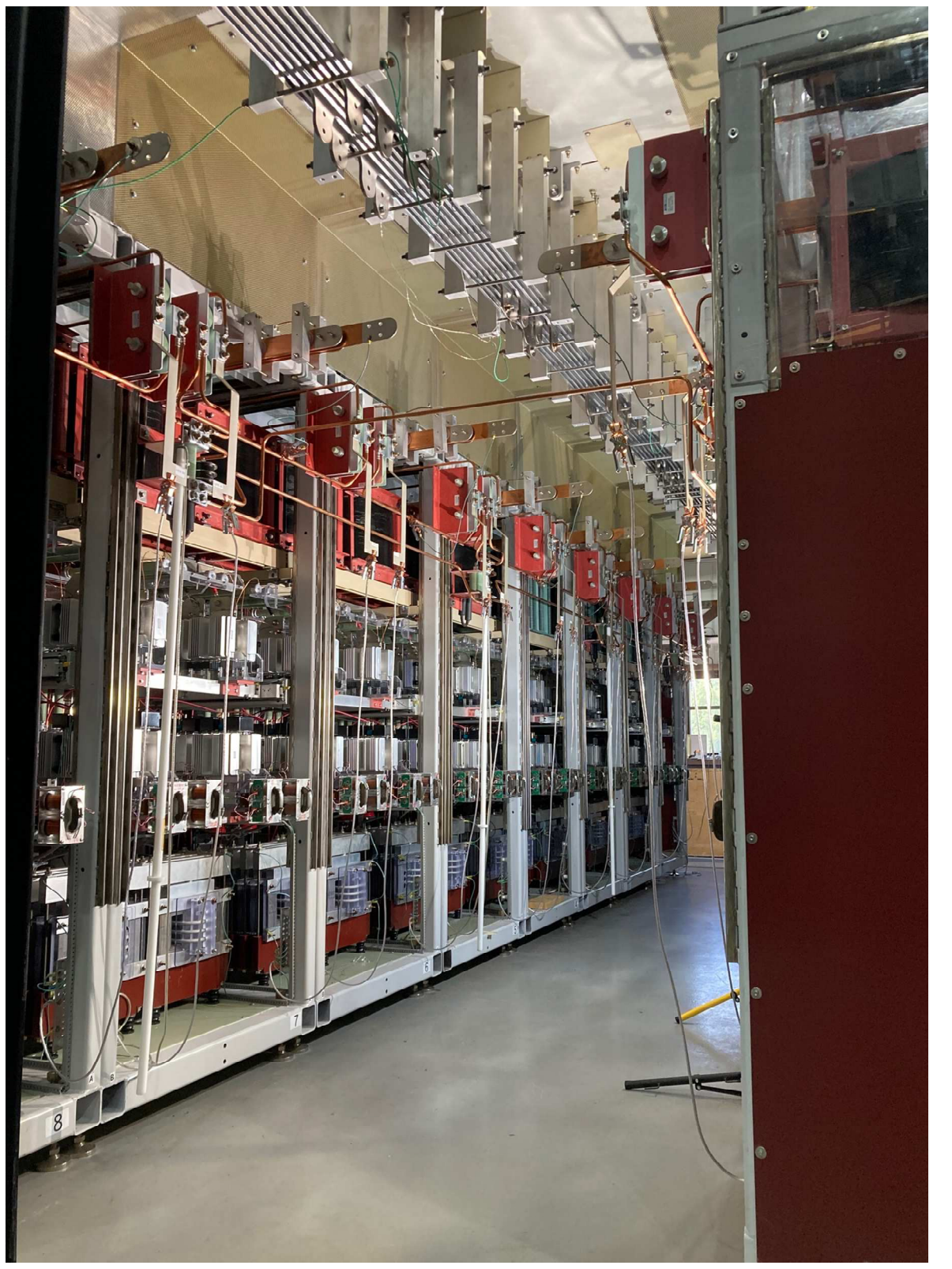

Fig. 4. Finished Assembly of Pulser Units and Stripline

This ratio is very conservative but based on operating experience with similar stripline in a high radiation environment.

The connections between sections are a nickel plated lapped joint with an rms current density of 150 Arms per square inch and clamping force of 28,000 lbs. A variation in the thickness of the connections has caused a reduced area of current contact and uneven pressure under the clamp. An expanded metal foil made of tin plated copper has been tested at low current. It was found to make the contact area pressure more uniform and reduce electrical resistance. They will be installed in some joints to see how they work with a high pulsed current.

## V. Full System Test

All the cabinets have been installed in the final location after completing pulse testing into a dummy load. The stripline was then assembled and mounted to the cabinets as shown in Fig 4. Unfortunately, while the horn is complete it is not yet attached to the stripline so we cannot test yet. The horn cooling system is incomplete and is needed to run at full repetition rate but not at low repetition rate.

Commissioning will start with 16 of the 32 cells being powered to about 75% of full peak current, 75% pulse length and low repetition rate. Once we have solved any unanticipated issues we will move to testing the full 32 cells with the cooling system and run at full peak current and repetition rate. Horn qualification requires only 200,000 pulses per unit but needs to be done for each horn made and for several configurations.

## VI. Acknowledgements

This project had an initial conceptual design in 2015. It has had several requirement changes, several design cycles and changes and had many different people contributing. The following are recognized for their contribution to the implementation: Ty M. Omark, Sabri Shawar, Noah M. Curfman, Gregory Meyer, David Durando, Mike Henry and John Anderson.

## VII. References

[1] C. C. Jensen, T. Omark, H. Pfeffer, K. Roon, J. Hugyik and B. Morris, "A 300 kA Pulsed Power Supply for LBNF," 2022 IEEE International Power Modulator and High Voltage Conference (IPMHVC), Knoxville, TN, USA, 2022, pp. 1-4

[2] H. Pfeffer, M. Davidson, N. Curfman and T. Omark, "Design Challenges in High-Current Pulsed Striplines," 2022 IEEE International Power Modulator and High Voltage Conference (IPMHVC), Knoxville, TN, USA, 2022, pp. 45-48

[3] K. Bourkland, K. Roon and D. Tinsley, "205 kA pulse power supply for neutrino focusing horns," Conference Record of the Twenty-Fifth International Power Modulator Symposium, 2002 and 2002 High-Voltage Workshop., Hollywood, CA, USA, 2002, pp. 266-269